\documentclass[12pt]{article}
\usepackage{amsxtra,amssymb,amsthm,amsmath,latexsym}

\theoremstyle{plain}

\newtheorem{theorem}{Theorem}[section]

\newtheorem{remark}[theorem]{Remark}

\def\ds{\displaystyle}

\def\R{{\mathbb R}}

\def\C{{\mathbb C}}
\def\oH{\buildrel\circ\over H}
\def\oH1{\buildrel\circ\over H\kern-.02in{}^1}

\begin{document}


\title{ An inverse problem of ocean acoustics
   \thanks{Math subject classification: 35R30 }
}

\author{
A.G. Ramm\\
 Mathematics Department, Kansas State University, \\
 Manhattan, KS 66506-2602, USA\\
ramm@math.ksu.edu\\
}

\date{}

\maketitle\thispagestyle{empty}

\begin{abstract}
Let
\begin{equation}
 \Delta u+k^2n(z)u=-\frac{\delta(r)}{2\pi r} f(z)
 \hbox{\ in\ } \R^2\times [0,1],
\tag{1}\end{equation}
\begin{equation}
u(x^1,0)=0, \quad u^\prime (x^1,1)=0,  
\tag{2}\end{equation}
where $u=u(x^1,z),\,
 \quad x^1 := (x_1, x_2),\,  r:= |x^1|,
\quad x_3 := z, \quad u^\prime = \frac{\partial u}{\partial z},$
$\delta (r)$ is the delta-function, 
$n(z)$ is the refraction coefficient,
which is assumed to be a real-valued integrable function,
$k>0$ is a fixed wavenumber. 
The solution to (1)-(2) is selected by the
limiting absorption principle.

It is proved that if $f(z)= \delta (z-1)$, then $n(z)$ is uniquely
determined by the data $u(x^1, 1)$ known $ \forall x^1 \in
\R^2$. Comments are made concerning the earlier study
of a similar problem in the literature.

\end{abstract}


\section{Introduction}

In~\cite{1} the following inverse problem is studied:

\begin{equation}
[\Delta +k^2n(z)]u = -\frac{\delta(r)}{2 \pi r} f(z), \quad
\hbox{\ in\ }
\R^2 \times [0,1],
\tag{1.1}\end{equation}

\begin{equation}
u(x^1,0)= u^\prime (x^1,1)=0, \quad x^1 :=
(x_1,x_2), \quad x_3 :=z, \quad
u^\prime := \frac{\partial u}{\partial z}.
\tag{1.2}\end{equation}

Here $k>0$ is a fixed
 wavenumber, $n(z)>0$ is the refraction coefficient,
which is assumed in [1] to be a continuous real-valued
function satisfying the condition $0\leq n(z)<1$, the
layer $\R^2 \times [0,1]$ models shallow ocean,
$r := |x^1| = \sqrt{x_1^2 + x_2^2}, \quad \delta (r)$
is the delta-function, $\frac {\delta(r)}{2\pi r}=\delta(x^1)$,
 $f(z) \in C^2 [0,1]$ is a 
function satisfying the
following conditions~\cite{1}, p.127:
$$f(0) = f^{\prime \prime} (0) = f^\prime (1)=0, \quad
f^\prime (0) \neq 0, \quad  f(1) \neq 0, \quad f(z) >0\,\, 
\hbox{\ in\ }\,\, 
(0,1). \quad  (C)
$$

The solution to (1.1)-(1.2) in~\cite{1} is 
required to satisfy some conditions
(~\cite{1}, p. 122, formulas~(1.4),
(1.8)-(1.10)) of the radiation conditions type. 

It is convenient to define the solution as $u(x) =
\ds \lim_{\varepsilon \downarrow 0} u_{\varepsilon}(x)$,
that is by the limiting absorption principle.
We do not show the dependence on $k$ in $u(x)$ since
$k>0$ is fixed throughout the paper.
The function $ u_{\varepsilon}(x)$ is the unique solution to
problem (1.1)--(1.2) in which equation (1.1) is replaced by
the equation with absorption:
$$
[\Delta +k^2n(z)-i\varepsilon]u_{\varepsilon}(x) =
-\frac{\delta(r)}{2 \pi r}f(z), \quad
\hbox{\ in\ }
\R^2 \times [0,1],\,\,\varepsilon >0.
$$
One defines the differential operator corresponding to 
differential expression (1.1)
and the boundary conditions (1.2) in $L^2(\R^2 \times [0,1])$
as a selfadjoint operator (for example, as
the Friedrichs extension of the symmetric operator
with the domain consisting of $H^2(\R^2 \times [0,1])$
functions vanishing near infinity and satisfying conditions
(1.2)), and then the function 
$u_{\varepsilon}(x)$ is uniquely defined.
By $H^m$ we mean the usual Sobolev space. One can prove that
the limit of this function  $u(x) =
\ds \lim_{\varepsilon \downarrow 0} u_{\varepsilon}(x)$
does exist globally in the weighted space 
$L^2(\R^2 \times [0,1],\frac 1 {(1+r)^a}), \, a>1,$
and locally in $H^2(\R^2 \times [0,1])$ outside
a neighborhood of the set $\{r=0,\, 0\leq z \leq 1\}$,
provided $\lambda_j \neq 0\,\, \forall j$, where $\lambda_j$ are
defined in (1.7) below. This limit
defines the unique solution to problem (1.1)--(1.2)
satisfying the limiting absorption principle
if $\lambda_j \neq 0 \,\, \forall j$. 
If $f(z)=\delta(z-1)$, where $\delta(z-1)$ is the 
delta-function, then an analytical
formula for $u_{\varepsilon}(x)$ can be written:
$$
u_{\varepsilon}(x)=\sum_{j=1}^{\infty}\psi_j(z)f_j
\frac 1{2\pi}K_0(r\sqrt{\lambda_j^2+i\varepsilon}),
$$ 
where $K_0(r)$ is the modified Bessel function
(the Macdonald function), and $f_j=\psi_j(1)$ are
defined in (1.6) below, and $\psi_j(z)$ and $\lambda_j^2$
are defined in formula (1.7) below.
This formula can be checked by direct calculation
and is obtained by the separation of variables.
The known formula $\mathcal F^{-1} \frac 1 {\lambda^2 +a^2}=
\frac 1 {2\pi}K_0(ar)$ was used, and $\mathcal F u:=\hat u$ is the Fourier
transform defined above formula (1.3).

From the formula for $u_{\varepsilon}(x)$, the known asymptotics  
$K_0(r)=\sqrt{\frac {\pi}{2r}}e^{-r}[1+O(r^{-1})]$
for large values of $r$, the boundedness of $|\psi_j(z)|$
as $j\to \infty$ and formula (1.8) below,
one can see that the limit of  $u_{\varepsilon}(x)$
as $\varepsilon \to 0$ does exist for any $r>0$ and $z\in
[0,1]$, if and only if $\lambda_j \neq 0.$ If $\lambda_j=0$ for some
$j=j_0$,
then the limiting absorption principle holds if and only if
$f_{j_0}=0$. If  $\lambda_j \neq 0 \,\, \forall j$, 
then the limiting absorption principle holds and the solution
to problem (1.1)-(1.2) is well defined.
If  $\lambda_j=0$ for some
$j=j_0$, then we define the solution to problem (1.1)-(1.2) with
$f(z)=\delta(z-1)$ by the formula:
$$
u(x)=\psi_{j_0}(z)\psi_{j_0}(1)\frac 1{2\pi}
\log(\frac 1r) +
\sum_{j=1,\, j\neq j_0}^{\infty}\psi_j(z)\psi_j(1)
\frac 1{2\pi}K_0(r\lambda_j),\quad r:=|x^1|.
$$ 
This solution is unique in the class of functions of the form
$u(x)=\sum_{j=1}^{\infty}u_j(x^1)\psi_j(z)$, where $\Delta_1 u_j-
\lambda_j^2 u_j=-\delta (x^1)$ in $\R^2$, $\Delta_1 w:= w_{x_1x_1}+
w_{x_2x_2}$, $u_j \in L^2(\R^2)$ if $\lambda_j^2 >0$;
if  $\lambda_j^2 <0$ then
 $u_j $ satisfies the radiation condition 
$r^{1/2}(\frac {\partial u_j}{\partial r}-i|\lambda_j|u_j) \to 0$
as $r\to \infty$, uniformly in directions $\frac {x^1}{r}$;
and  if $\lambda_j^2 =0$ then $u_j=\frac 1{2\pi}
\log(\frac 1r) +o(1)$ as $r\to \infty$.
  
{\it The inverse problem (IP) consists of finding
$n(z)$ given $g(x^1) := u(x^1,1)$ and assuming that
$f(z)=\delta(z-1)$ in (1.1).}

 By the cylindrical symmetry one has
$g(x^1) = g(r)$.

It is claimed in~\cite[p.~137]{1} that the above inverse problem
has not more than one solution, and a method for 
finding this solution
is proposed. The arguments in \cite{1} are not satisfactory
(see Remark 2.1 below, where some of the incorrect 
statements from [1],
which invalidate the approach in [1],
are pointed out).

{\it The aim of our paper is to prove that
if $f(z)= \delta(z-1)$, then $n(z)$ can be uniquely 
and constructively determined
from the data $g(r)$ known for all $r>0$.}
It is an open problem to find all such $f(z)$ for
which the $IP$ has at most one solution.

The method we use is developed 
in~\cite{5} (see also \cite{7}). Properties of the operator
$\Delta +k^2n(z)$ in a layer were studied in~\cite{6}. In~\cite{8}
an inverse problem for an inhomogeneous Schr\"odinger equation
on the full axis was investigated. 

{\it Let us outline our approach to IP}. 

Take the
Fourier transorm of (1.1)-(1.2) with respect to $x^1$ and let
$$v := v (z,\lambda) :=\hat u:= \int_{\R^2} u(x^1, z) 
e^{ix^1\cdot \zeta} dx^1,
\quad |\zeta| := \lambda, \quad \zeta \in \R^2,$$
and
 $$G(\lambda) :=\hat g(r).$$

Then
\begin{equation}
\ell
v := v^{\prime \prime} - \lambda^2 v+q(z)v = -f(z), 
\quad q(z) := k^2n(z), \quad v=v(z,\lambda),
\tag{1.3}\end{equation}
\begin{equation}
v(0, \lambda) = v' (1,\lambda) = 0,
\tag{1.4}\end{equation}
\begin{equation}
v(1, \lambda) = G(\lambda).
\tag{1.5}\end{equation}

{\it IP: The inverse problem is: given $G(\lambda)$, for all
$\lambda >0$ and a fixed
$f(z)=\delta(z-1)$,
find $q(z)$.}

The solution to (1.3)-(1.4) is:
\begin{equation}
v(z,\lambda) = \sum^\infty_{j=1} \frac{\psi_j(z) f_j}
{\lambda^2 + \lambda_j^2}, \quad f_j := (f,\psi_j) := \int^1_0
f(z) \psi_j(z) dz,
\tag{1.6}\end{equation}
where $\psi _j (z)$ are the real-valued normalized eigenfunctions of the
operator
$L := -\frac{d^2}{dz^2} - q(z)$:
\begin{equation}
L \psi_j = \lambda_j^2 \psi_j, \quad \psi_j (0) = \psi^\prime_j
(1) =0, \quad ||\psi_j(z)||=1.
\tag{1.7}\end{equation}
We can choose the eigenfunctions $\psi_j(z)$ real-valued since
the function $q(z)=k^2n(z)$ is assumed real-valued. 
One can check that all the eigenvalues are simple, that is, there is 
just one eigenfunction $\psi_j$ corresponding to the eigenvalue
$\lambda_j^2$ (up to a constant factor, which for
real-valued normalized eigenfunctions can be either $1$ or $-1$).

It is known (see e.g. [4. p.71]) that
\begin{equation}
\lambda_j^2 = \pi^2 (j-\frac{1}{2})^2 [1+O(\frac 1{j^2})] \hbox{\ as\ } j
\to +\infty.
\tag{1.8}\end{equation}

The data can be written as
\begin{equation}
G(\lambda) = \sum^\infty_{j=1} \frac{\psi_j(1) f_j}
{\lambda^2 + \lambda_j^2},
\tag{1.9}\end{equation}
where $f_j$ are defined in (1.6).
The series (1.9) converges absolutely and uniformly
on compact sets of the complex plane $\lambda$ outside 
the union of small discs centered at the points $\pm i\lambda_j$.
Thus, $G(\lambda)$ is a meromorphic function on the whole
complex $\lambda$-plane with simple poles at the points
$\pm i\lambda_j$. Its residue at $\lambda= i\lambda_j$
equals $\frac {\psi_j(1)f_j}{2i\lambda_j}$.

If $f(z) = \delta(z-1)$, then $f_j = \psi_j (1) \neq 0\,\, 
\forall j=1,2,.....$, 
(see section 2 for a proof of the inequality 
$\psi_j (1) \neq 0 \,\, 
\forall j=1,2,.....$,)
and the data (1.9)
determine uniquely the set
\begin{equation}
\{\lambda_j^2, \quad \psi_j^2 (1)\}_{j=1,2, \dots}
\tag{1.10}\end{equation}

In section 2 we prove the basic result:
\begin{theorem}
If $f(z) = \delta(z -1)$ then the data (1.5) 
determine $q(z) \in L^1 (0,1)$
uniquely.
\end{theorem}
An algorithm for calculation of $q(z)$ from the data is
described in section 2.
 
\begin{remark}  The proof and the conclusion
of Theorem 1.1 remain valid for other boundary conditions,
for example, $u'(x^1,0)=u(x^1,1)=0$ with the data $u(x^1,0)$
known for all $x^1\in \R^2$.

\end{remark}

\section{Proofs: uniqueness theorem and inversion algorithm}

\begin{proof}[Proof of Theorem 1.1]
The data (1.9) with $f(z)=\delta(z-1)$, that is, with
 $f_j=\psi_j(1)$, determine uniquely
$\{\lambda_j^2\}_{j = 1,2, \dots}$ since $\pm i\lambda_j$ are 
the poles of the meromorphic function $G(\lambda)$ which is uniquely
determined for all $\lambda \in \C$ by its values for all $\lambda>0$
(in fact, by its values at any infinite sequence of $\lambda >0$ which
has a finite limit point on the real axis).
The residues
$\psi_j^2 (1) \hbox{\ of\ } G(\lambda) \hbox{\ at\ } \lambda =
i\lambda_j$
are also uniquely determined.

Let us show that:
\begin{description}
\item{i)} $\psi_j (1) \neq 0 \quad \forall j= 1,2,\dots$

\item{ii)} The set (1.10) determines $q (z) \in L^1(0,1)$ uniquely.
\end{description}

{\it Let us prove i)}: 

If $\psi_j (1) =0$ then equation (1.7) and the Cauchy
data $\psi_j (1) = \psi_j^\prime (1) =0$
imply that $\psi_j(z) \equiv 0$ which is impossible since
$\parallel \psi_j(z) \parallel =1$, 
where $ \parallel u \parallel^2 :=
\int^1_0 |u|^2dx$.

{\it Let us prove ii):} 

It is sufficient to prove that the set
(1.10) determines the norming constants
$$\alpha_j:=\parallel \Psi_j(z) \parallel^2$$ 
and therefore the set
$$\{ \lambda_j^2, \alpha_j\}_{j=1,2, \dots},$$
where the eigenvalues $\lambda_j^2$ are defined
in (1.7),  $\Psi_j = \Psi(z,\lambda_j),\, \psi_j(z):=
\frac {\Psi(z,\lambda_j)}{\parallel \Psi_j \parallel}$, 
\begin{equation}
-\Psi^{\prime \prime} -s^2 \Psi -q(z) \Psi =0, 
\quad \Psi(0,s) = 0,\quad \Psi'(0,s) = 1,
\tag{2.1}\end{equation}
and $\lambda_j$ are the zeros of the equation
\begin{equation}
\Psi' (1,s) = 0, \quad s=\lambda_j, \,\, j=1,2,.......
\tag{2.2}\end{equation}

The function $\Psi'(1,s)$ is an entire function of $\nu=s^2$
of order
$\frac{1}{2}$, so that (see~\cite{2}):
\begin{equation}
\Psi' (1,s) =\gamma \prod^\infty_{j=1}
\left(1-\frac{s^2}{\lambda_j^2}\right),
 \quad \gamma=const.
\tag{2.3}\end{equation}
From the Hadamard factorization theorem for entire functions
of order $<1$ formula (2.3) follows but the constant factor
$\gamma$ remains undetermined.
This factor is determined by the data $\{\lambda_j^2\}_{\forall j}$
 because the main term of the asymptotics
of function (2.3) for large positive $s$ is $\cos (s)$, and
the result in \cite{4}, p.243, (see Claim 1 below) implies that the
constant $\gamma$ in formula (2.3) can be computed explicitly:
\begin{equation}
\gamma= \prod^\infty_{j=1}\frac {\lambda_j^2}{(\lambda_j^0)^2},
\tag{2.3'}\end{equation}
where $\lambda_j^0$ are the roots of the equation $\cos(s)=0$,
$\lambda_j^0=\frac {(2j-1)\pi}2,\,\,j=1,2,.....$, and the 
infinite product in (2.3') converges because of (1.8). 

A simple derivation of (2.3'), independent of the result formulated in
Claim 1 below, is based on the formula:
$$
1=\lim_{y\to +\infty} \frac {\Psi'(1,iy)}{\cos(iy)}=\gamma
\prod_{j=1}^\infty \frac {(\lambda_j^0)^2}{\lambda_j^2}.
$$
 
For convenience of the reader let us formulate the result
from \cite{4}, p.243, which yields formula (2.3') as well:

{\it Claim 1: The function $w(\lambda)$ admits the
representation
$$w(\lambda)=\cos(\lambda) -B\frac {\sin (\lambda)}{\lambda} +\frac
{h(\lambda)}{\lambda},$$
where $B=const,$ $h(\lambda)=\int_0^1 H(t)\sin(\lambda t) dt,$ 
and $H(t)\in L^2(0,1)$ if and only if
$$w(\lambda)=\prod^\infty_{j=1}\frac
{\lambda_j^2 -\lambda^2}{(\lambda_j^0)^2},$$ 
where $\lambda_j=\lambda_j^0 -\frac B j+ \frac {\beta_j}{j},$
$ \beta_j$ are some numbers satisfying the condition:
$\sum_{j=1}^{\infty}|\beta_j|^2<\infty,$   $\lambda_j$
are the roots of the even
function $w(\lambda)$ and $\lambda_j^0=
(j-\frac 12)\pi,\,\,j=1,2,.....,$ are the 
positive roots of $\cos(\lambda)$.}

The equality
\begin{equation}
\prod^\infty_{j=1}\frac
{\lambda_j^2 -\lambda^2}{(\lambda_j^0)^2}=\gamma \prod^\infty_{j=1}
\left(1- \frac{\lambda^2}{\lambda_j^2}\right),
\tag{2.3"}\end{equation}
where $\gamma$ is defined in (2.3'), is easy to prove:
if $w$ is the left-hand
side and $v$ the right-hand side of the above equality, then
 $w$ and $v$ are entire functions of $\lambda$,
the infinite products converge absolutely,
$\frac
{\lambda_j^2 -\lambda^2}{(\lambda_j^0)^2}=\frac {\lambda_j^2}{(\lambda_j^0)^2}
\left (1- \frac{\lambda^2}{\lambda_j^2}\right),$ and
 taking the infinite product and using 
(2.3'), one concludes that $\frac w v=1$, as claimed.

In fact, one  can establish formula (2.3") 
and prove that $\gamma$ in (2.3") is defined by (2.3') without
assuming a priori that (2.3') holds and without using Claim 1.
The following assumption suffices for the proof of (2.3"):

i) $\lambda_j^2=(\lambda_j^0)^2+ O(1), \,\, 
(\lambda_j^0)^2=\pi^2(j-\frac 12)^2.$ 

Indeed, if i) holds then both sides of (2.3") are entire functions
with the same set of zeros and their ratio is 
a constant. This constant equals to $1$ if there is a sequence of points
at which this ratio converges to $1$. 
Using the known formula:
$\cos (\lambda)=\prod^\infty_{j=1}\frac
{(\lambda_j^0)^2 -\lambda^2}{(\lambda_j^0)^2} $, and the assumption i)
one checks easily that the ratio of the left- and right-hand sides of
(2.3") tends to $1$ along the positive imaginary semiaxis. 
Thus, we have proved formulas (2.3)-(2.3') without reference to Claim 1.

 The above claim is used with $w(s)=\Psi'(1,s)$ in our paper.
The fact that $\Psi'(1,s)$ admits the representation
required in the claim is checked by means of the formula
for $\Psi'(1,s)$ in terms of the transformation operator:
$\Psi(z,s)=\frac {\sin (sz)}{s}+
\int_0^z K(z,t)\frac {\sin (s t)}{s}dt$,
and the properties of the kernel $K(z,t)$ are studied in
\cite {4}. Thus, 
$\Psi'(1,s)=\cos(s) +\frac {K(1,1)\sin(s)}
{s} +\int_0^1K_z(1,t)\frac {\sin (s t)}{s}dt$. This
is the representation of  $\Psi'(1,s):=w(s)$
used in  Claim  1.

Let us derive a formula for $\alpha_j:=\parallel \Psi_j \parallel^2$.
Denote $\dot\Psi := \frac{d \Psi}{d \nu}$, differentiate
(2.1), with $s^2$ replaced by $\nu$,
with respect to $\nu$ and get:
\begin{equation}
-\dot\Psi^{\prime \prime} - \nu \dot\Psi - q \dot\Psi = \Psi.
\tag{2.4}\end{equation}

Since $q(z)$ is assumed real-valued, 
one may assume $\psi$ real-valued.
Multiply (2.4) by $\Psi$ and (2.1) by $\dot\Psi$, 
subtract and integrate
over $(0,1)$ to get
\begin{equation}
0< \alpha_j:=\int^1_0 \Psi_j^2 dz =
\left(\Psi_j' \dot\Psi_j - \Psi_j \dot\Psi'_j\right)
\bigg|_0^1 =- \Psi_j (1) \dot\Psi_j' (1),
\tag{2.5}\end{equation}
where the boundary conditions
$\Psi_j (0) = \Psi'_j (1) = \dot\Psi_j (0) =0$ were used.

From (2.3) with $s^2=\nu$ one finds the numbers $b_j:=\dot\Psi_j'(1)$:
\begin{equation}
b_j= \gamma \frac{d}{d \nu} \prod^\infty_{j'= 1}
\left( 1-\frac{\nu}{\lambda^2_{j'}}\right)
\bigg|_{\nu = \lambda_j^2}
= -\frac{\gamma}{\lambda_j^2} \prod_{j^\prime \neq j}
\left(1-\frac{\lambda^2_j}{\lambda_{j'}^2}\right).
\tag{2.6}\end{equation}

{\it Claim 2: The data
$\psi_j^2 (1)=\frac {\Psi_j^2 (1)}{\alpha_j}:=t_j $,
where $\alpha_j:=\parallel \Psi_j(z)\parallel^2$, and 
equation (2.5) determine uniquely $\alpha_j$.} 

Indeed, the numbers $b_j$ are the known numbers from
formula (2.6). Denote by $t_j:=\psi_j^2 (1)$ the quantities known
from the data (1.10). Then
it follows from (2.5) that $\alpha_j^2=t_j \alpha_j b_j^2$,
so that 
\begin{equation}
\alpha_j=t_jb_j^2.
\tag{2.7}
\end{equation}
Claim 2 is proved. 

Thus, the data (1.10) determine $\alpha_j=\parallel \Psi_j \parallel^2$
uniquely and analytically by the above formula,
and consequently $q(z)$ is uniquely determined by the
following known theorem
(see for example, [3]):

{\it The spectral function of the operator $L$ 
determines $q(z)$ uniquely.}

The spectral function $\rho(\lambda)$ of
the operator $L$ is defined by the formula (see [3, formula (10.5)]):
\begin{equation}
\rho (\lambda) = \sum_{\lambda_j^2 < \lambda}
\frac{1}{\alpha_j}.
\tag{2.8}\end{equation}

The Gelfand-Levitan algorithm [3] allows one to reconstruct 
analytically
$q(z)$ from the spectral function $\rho (\lambda)$
and therefore from the data (1.10), since, as we have proved already,
these data determine the spectral function $\rho(\lambda)$ uniquely.

Theorem 1.1 is proved. 
\end{proof}

{\it Let us describe an algorithm for calculation of $q(z)$ from the data
$g(x^1)$:} 

{\it Step 1}: Calculate $G(\lambda)$, the Fourier transform of
$g(x^1)$.
Given $G(\lambda)$, find its poles $\pm i\lambda_j$, and 
consequently the numbers $\lambda_j$; then
find its residues, and consequently the numbers $\psi_j(1)f_j$.

{\it Step 2}: Calculate the function (2.3), and the constant
$\gamma$ by formulas (2.3) and (2.3'). Calculate 
the numbers $b_j$ by formula (2.6) and $\alpha_j$
by formula (2.7). Calculate the spectral function 
$\rho (\lambda)$ by formula (2.8).

{\it Step 3}: Use the known Gel'fand-Levitan algorithm (see
[3]-[5]) to calculate $q(z)$ from $\rho (\lambda)$.

{\it This completes the description of the inversion algorithm
for IP.}

\begin{remark}
There are inaccuracies in [1]. We point out two
of these, of which the first invalidates the approach in [1].

In [1, p.128, line 2] the $\alpha_n$ are not the same as
$\alpha_n$ in formula [1, (3.3)]. If one uses $\alpha_n$
from formula [ 1, (3.3)], then one has to use in [1, p.128, line 2]
the coefficients $\alpha_n \phi_n(h)$, according to
formula [1, (1.5)]. In [1] $h$ is the width of the layer,
which we took to be $h=1$ in our paper without loss of generality.
However, the numbers $\phi_n(h)$ are not known
in the inverse problem, since the coefficient $n(z)$ is not
known. Therefore formula [1, (3.9)] is incorrect. This invalidates
the approach in [1]. 

In [1, p.128] a negative decreasing sequence 
of real numbers $a_n$ is
defined by equation (3.1), which we give for $h=1$:
$$k \sqrt {1-a_n^2}=(n+\frac 12)\pi +O(\frac 1n) \quad (\ast).$$
 Such a sequence does not
exist: if $a_n<0$ and $a_n$ has a 
finite limit then the right-hand side of
$(\ast)$ cannot grow to infinity, 
and if $a_n \to -\infty$, then the left-hand side of $(\ast)$ cannot 
stay positive for large $n$, and therefore
cannot be equal to the right-hand side of $(\ast)$.

\end{remark}

\end{document}